\documentclass[12pt]{article}
\usepackage{amsmath,amssymb}
\usepackage{epsfig} 

\newcommand{\newc}{\newcommand}
\newc{\gev}{\hbox{\rm\,GeV}}
\newc{\tev}{\hbox{\rm\,TeV}}

\newc{\gsim}{\lower.7ex\hbox{$\;\stackrel{\textstyle>}{\sim}\;$}}
\newc{\lsim}{\lower.7ex\hbox{$\;\stackrel{\textstyle<}{\sim}\;$}}

\def\simlt{\stackrel{<}{{}_\sim}}

\def\eq#1{eq.~(\ref{#1})}

\def\vev#1{\langle {#1} \rangle}
\def\beq{\begin{equation}}
\def\eeq{\end{equation}}
\def\bea{\begin{eqnarray}}
\def\eea{\end{eqnarray}}

\def\gsm{g_{\scriptscriptstyle {SM}}}

\begin{document}

\baselineskip=18pt

\setcounter{footnote}{0}
\setcounter{figure}{0}
\setcounter{table}{0}

\begin{titlepage}
\begin{flushright}
CERN-PH-TH/2007--234\\
\end{flushright}
\vspace{.3in}

\vspace{.5cm}

\begin{center}

{\Large\sc{\bf Natural $\mu$ and $B\mu$ in gauge mediation}}

\vspace*{9mm}
\renewcommand{\thefootnote}{\arabic{footnote}}

\mbox{ \bf Gian F.~Giudice$^{\,1}$, Hyung Do Kim$^{\,2,3}$, Riccardo Rattazzi$^{\,2}$}

\vspace*{0.9cm}

{\it $^1$ CERN, Theory Division,  CH--1211 Geneva 23, Switzerland.}

{\it $^2$ Institut de Th\'eorie des Ph\'enom\`enes Physiques, EPFL, CH--1015 Lausanne, Switzerland.}

{\it $^3$ FPRD and Department of Physics, SNU, Seoul, Korea.}

\end{center}

\vspace{1cm}

\begin{abstract}
\medskip
\noindent
We propose a natural solution to the $\mu$ problem in gauge mediation. It relies on the logarithmic dependence of the effective  K\"ahler potential on the messenger threshold superfield $X$. Thus, $\mu$ and $B\mu$ naturally arise at one and two loops, respectively. Moreover $B$ has the same phase as  the gaugino mass and   the supersymmetric CP problem is solved as well.
\end{abstract}

\bigskip
\bigskip

\end{titlepage}


\section{Introduction} 
\label{sec1}

Gauge mediation~\cite{dine1}--\cite{grrev} is an attractive realization of low-energy supersymmetry which successfully explains the absence of large flavor violations. Its main difficulty lies in the generation of proper values for the higgsino mass $\mu$ and the Higgs mass mixing $B\mu$. Indeed, once a mechanism for generating $\mu$ is found, one generically obtains the relation~\cite{dynmu}
\beq
B = \frac {B\mu}{\mu} \simeq \frac FM,
\label{prob}
\eeq
where $\sqrt{F}$ is the supersymmetry-breaking scale and $M$ is the messenger mass. Since soft terms are characterized by the scale ${\tilde m} \sim g_{\scriptscriptstyle {SM}}^2F/(16\pi^2M)$, where $g_{\scriptscriptstyle {SM}}$ collectively denotes the gauge couplings, \eq{prob} gives the phenomenologically unacceptable prediction that $B$ is two orders of magnitude larger than ${\tilde m}$. This $\mu$($B$) problem is a characteristic of all theories in which the soft terms are derived from the original scale of supersymmetry breaking through small parameters, and it is absent in theories like gravity mediation~\cite{giumas}.

This problem cannot be ignored in any realistic construction. Indeed, it is rather pointless to build models of gauge mediation without addressing the $\mu$($B$) problem. After all, the main motivation of low-energy supersymmetry is to produce a plausible and realistic theory of electroweak breaking. This cannot be achieved if $\mu$ and $B$ are not of the size of the other soft terms. Therefore, if we want to derive meaningful phenomenological predictions or to assess the relative merit of different schemes of supersymmetry-breaking mediation, we should consider only models of gauge mediation with a proper mechanism for $\mu$ and $B\mu$.

So far, three kinds of solutions to the $\mu$($B$) problem in gauge mediation have been proposed. The first~\cite{dynmu} is to generate $\mu$ at one loop through the $D$ term of a higher covariant-derivative effective operator. Such an operator does not generate $B\mu$, which is induced only at the next order in perturbation theory.
The second class of solutions is based on a new weak-scale
singlet superfield $S$ coupled to the Higgs bilinear in the
superpotential. The correct pattern of gauge symmetry breaking can be obtained if one extends the minimal model to include appropriate couplings between $S$ and the messengers~\cite{grrev,sla} (see also ref.~\cite{chac}), or non-renormalizable couplings of $S$~\cite{dine3}, or additional light fields~\cite{dine1} (see also ref.~\cite{mura}). Finally, it was recently suggested~\cite{conf} that strongly-interacting dynamics in the hidden sector can efficiently suppress the dimension-two soft parameter $B\mu$  with respect to the dimension-one parameter $\mu$, in the renormalization from high to low energies, thus solving the $\mu$($B$) problem. In this mechanism, the characteristic mass spectrum of gauge mediation in the squark and slepton sector is completely obliterated. 
In this paper, we want to propose a new solution to the $\mu$($B$) problem in gauge mediation. 

\section{The mechanism}
\label{sec2}

To have one-loop generated $\mu$, but not to $B\mu$, it is necessary that the effective action, after integrating out the messengers at one loop, be of the form
\beq
\int  d^4\theta H_uH_d \left[ f(X) + g(X^\dagger) + D^2 h(X,X^\dagger) \right] + {\rm h.c.}
\label{opmu}
\eeq
Here $D_\alpha$ is the supersymmetric covariant derivative and $f,g,h$ are generic functions of the hidden-sector chiral superfield $X$ containing the Goldstino, with background value $X=M+\theta^2F$.
The mechanism proposed in ref.~\cite{dynmu} relies on the third term in \eq{opmu}. Here we want to exploit the case in which the dependence on $X$ splits into the sum of holomorphic and anti-holomorphic functions, and use the second term in \eq{opmu} to generate $\mu$. No $B\mu$ is induced at the one-loop level\footnote{This possibility was also commented in footnotes in refs.~\cite{grrev,ibe}, but no dynamical mechanism was proposed.}.

This problem has a close analogy with the generation of soft scalar squared masses ${\tilde m}_Q^2$. It is well known that in gauge mediation there is no one-loop contribution to ${\tilde m}_Q^2$, as a consequence of two essential ingredients of the theory. The first is a chiral reparametrization $U(1)_X$ invariance $X\to e^{i\varphi}X$,
with messenger fields transforming as ${\bar \Phi} \Phi \to e^{-i\varphi}{\bar \Phi} \Phi$. The second ingredient consists in having a messenger mass threshold fully determined by the $X$ superfield (indeed the mass term is $X{\bar \Phi} \Phi$). From these two properties we infer that the one-loop renormalization for the kinetic term of the matter superfield $Q$ must be of the form
\beq
 \int  d^4\theta  \left( 1+\frac{g^2}{16\pi^2} \ln \frac{X^\dagger X}{\Lambda^2}\right) Q^\dagger Q,
 \label{opq}
 \eeq
 where $\Lambda$ is the ultraviolet cutoff and $g$ some coupling constant.
 In the case of minimal gauge mediation, $g=0$ because matter is not directly coupled to the messenger sector. However, one loop-contributions are present in models with gauge messengers~\cite{gmm} or in models with direct matter-messenger couplings. In \eq{opq}, knowledge of the $\Lambda$ dependence (which is given by the supersymmetric RG equations) fully characterizes the structure of the soft terms~\cite{wave}. In particular, we observe that the $X$ dependence in \eq{opq} splits into the sum of a holomorphic and an anti-holomorphic part, and therefore no one-loop ${\tilde m}_Q^2$ is generated once we replace $X=M+\theta^2F$, although $A$ terms are induced.
 
This familiar result suggests a simple approach to address the $B$ problem of gauge mediation.
 Let us suppose that the ordinary (non-$R$) Peccei-Quinn (PQ) symmetry under which $H_uH_d$ has non-zero charge is broken,
 and yet no $\mu$-term appears in the superpotential. This property  may be enforced in a technically natural way thanks to the  non-renormalization theorem. It may also arise in a more natural way
 by assuming analyticity of the spurion that breaks PQ \cite{bns} or, in a fully natural way, by an additional $R$-symmetry under which $H_uH_d$ has charge  $\not =2$, for instance $[H_uH_d]_R=0$.  The last two cases
lead to  rather  plausible implementations in gravity mediation of the mechanism of ref.~\cite{giumas}. Let us also assume that the two essential ingredients of minimal gauge mediation are preserved:  $U(1)_X$ invariance and a messenger mass threshold fully characterized by $X$. Then, after the messengers have been integrated out, by power counting we should in general expect a one-loop contribution to the K\"ahler potential\footnote{In global supersymmetry the divergent term vanishes because $H_uH_d$ is holomorphic. However, this is not the case as soon as the Higgs is coupled to a non-trivial background, as in the case of supergravity where the presence of the superconformal compensator makes the operator non-holomorphic. This is analogous to the non-minimal gravitational coupling of a field to the Ricci scalar $\phi^2 R$, which is logarithmic divergent. In Minkowski background the divergence vanishes as $R=0$, but it is present in a curved background ($R\ne 0$).}
\beq
\int  d^4\theta \frac{g_{\rm eff}^2}{16\pi^2} \ln \frac{X^\dagger X}{\Lambda^2}~H_uH_d +{\rm h.c.},
\label{oplog}
\eeq
where $g_{\rm eff}^2$ indicates a combination of superpotential couplings.
This generates $\mu$ but not $B\mu$, which will be induced only at higher orders.

The difficulty with this approach is that the above result will never arise from a purely trilinear superpotential. This is because of the  presence of the ``trivial'' $R$-symmetry under which all fields, including $X$, carry charge $2/3$, thus implying   $g_{\rm eff}^2=0$. In order to explicitly break the trivial $R$-symmetry some dimensionful coupling must be introduced.
 By simple power counting,  $g_{\rm eff}^2$ must  be generated by the combined effect of super-renormalizable and non-renormalizable interactions. 
Then, in order to obtain a sizeable $\mu$, the ultraviolet cut-off associated with the non-renormalizable scale must be very close
to the other mass scales, a situation which is not very promising for model building.

However, this difficulty can be circumvented if the PQ symmetry is broken through a massive singlet superfield $S$ related to the Higgs bilinear $H_uH_d$ by its equation of motion. In this case, $R$-symmetry and renormalizability do not forbid the term $SM_1^\dagger \ln (X^\dagger X/\Lambda^2)$ in the K\"ahler potential, and the mechanism can go through. Here $M_1$ is a parameter related to the $S$ mass, which must be smaller than $M$, but can be much larger than the weak scale $\tilde m$.

To give a concrete example, let us consider one singlet superfield $S$ and two pairs of chiral messengers $\Phi =(\Phi_1,\Phi_2)$ and $\bar \Phi =({\bar \Phi}_1,{\bar \Phi}_2)$ with superpotential
\beq
W=\lambda SH_uH_d +\frac{M_2}{2} S^2 +\left( M_1 +\xi S \right) {\bar \Phi}_1\Phi_2 + X \left( {\bar \Phi}_1\Phi_1 + {\bar \Phi}_2\Phi_2 \right) .
\label{suppot1}
\eeq
Without loss of generality, we can take the coupling constants $\lambda$ and $\xi$ to be real. 
This model has a $U(1)_X$ invariance $X\to e^{i\varphi}X$ (with $\Phi_1$ and ${\bar \Phi}_2$ carrying charge $-1$)
and the messenger threshold is determined by $X$, if
we assume that the mass parameters $M_{1,2}$ are of comparable size, but much smaller than the messenger mass, $M_1\sim M_2 \ll M$. 

Integrating out the messenger fields at the scale $M$ generates a one-loop effective K\"ahler potential~\cite{grisa}
\beq
K_{\rm eff} = - \frac{1}{16\pi^2} \int d^4\theta ~{\rm Tr} \left( {\cal M}^\dagger {\cal M} \ln \frac{{\cal M}^\dagger {\cal M}}{\Lambda^2}\right)  .
\eeq
Here ${\cal M}$ is the (field-dependent) messenger mass matrix, defined as
\beq
W={\bar \Phi} {\cal M}  \Phi ,~~~~~{\cal M}=\left(\begin{array}{cc}
X &\epsilon \\ 0 & X \end{array}\right),~~~~~\epsilon \equiv M_1+\xi S .
\eeq
Computing the eigenvalues of ${\cal M}^\dagger {\cal M}$ and expanding in powers of $|\epsilon |/|X|$ (consistently with our assumption $M_{1} \ll M$), we find that the relevant terms in $K_{\rm eff}$ are given by
\bea
K_{\rm eff} &=& -\frac{5}{16\pi^2} \int d^4 \theta \left( \left| \epsilon \right|^2 \ln \frac{ \left| X \right|^2}{\Lambda^2} + \frac{\left| \epsilon \right|^4}{6\left| X \right|^2} + \dots  \right) 
= -\frac{5}{16\pi^2} \int d^4 \theta \Biggr[  \xi^2 S^\dagger S \ln \frac{X^\dagger X}{\Lambda^2}
 \nonumber \\ 
&+&\xi \left( M_1^\dagger S + {\rm h.c.}\right) \left( \ln \frac{X^\dagger X}{\Lambda^2} + \frac{M_1^\dagger M_1}{3 X^\dagger X}\right) 
+\frac{\xi^2 \left( M_1^{\dagger 2} S^2 + {\rm h.c.}\right)}{6X^\dagger X} +\dots \Biggr] .
\label{keff}
\eea
Here we have specified the case in which each $\Phi$ ($\bar \Phi$) fills a fundamental (anti-fundamental) representation of $SU(5)$. 

After replacing $X=M+\theta^2 F$, the log divergent term in \eq{keff} generates a superpotential linear in $S$ but no $S$ tadpole in the scalar potential, because of the special logarithmic functional dependence on $X^\dagger X$. Once we integrate out $S$ and use its equation of motion $S=-\lambda H_uH_d/M_2$, this term gives
\beq
\mu= 5\lambda \xi \frac{M_1^\dagger}{M_2}\left( \frac{F}{16\pi^2 M}\right)^\dagger .
\label{eccomu}
\eeq
By assuming  $M_1$ and $M_2$ have comparable size  and also $\lambda \sim\xi\sim \gsm$ we have $\mu
\sim {\tilde m} \sim \gsm^2 F/(16\pi^2 M)$. Since the log divergent term does not induce  an $S$ tadpole in the potential, there is no one-loop contribution to $B\mu$. Two-loop contributions are however  expected from double logarithmic renormalizations of the K\"ahler potential. Indeed, a simple calculation using the technique of ref.~\cite{wave} shows that\footnote{For simplicity we assume that the coupling $\xi$ is the same for the doublet and the triplet in the messenger multiplet. Also, we assume that $X$ is a non-propagating background field. These assumptions can be easily relaxed and do not alter the discussion. See ref.~\cite{sla} for general results.}
\beq
B =\left( \frac{16}{5} g_s^2+\frac 65 g^2 +\frac 23 g^{\prime 2}-2\xi^2 \right) \frac{F}{16\pi^2M},
\label{eccobmu}
\eeq
and therefore $B\mu$ is correctly predicted to be of order ${\tilde m}^2$.

On the other hand, the finite part of the linear term in $S$ in \eq{keff} generates an $S$ tadpole, giving a contribution
\beq
B = -\frac 13 \left| \frac{M_1}{M} \right|^2 \frac FM.
\eeq
Therefore, as long as we take $M_1/M \simlt g_{\scriptscriptstyle {SM}}/(4\pi)$, the finite contribution to $B\mu$ will be smaller than the two-loop effects and it can be neglected. 

From \eq{keff} we also infer that an $S^2$ term in the K\"ahler potential is only generated by finite contributions and therefore it is suppressed by $M_1^2/M^2$. This can be understood by considering a bookkeeping $R$-symmetry, where $S$ and $M_1$ carry the same charge. The term generated in the K\"ahler potential must be of the form $S^2 M_1^{\dagger 2}$ and therefore it is suppressed in the limit $M_1\ll M$.

This example illustrates how it is possible to generate a one-loop $\mu$ term, while  ensuring that no $B\mu$ term is induced at the same perturbative order. Notice that the low-energy theory at the weak scale has the usual field content of the minimal supersymmetric model. While messengers are integrated out at the scale $M$, the singlet $S$ has a mass $M_2$, and we are assuming $M\gg M_{1,2} \gg {\tilde m}$. 

The superpotential in \eq{suppot1}, which defines the example presented here, is non-generic, in the sense that it does not have the most general form consistent with symmetries. 
The addition of a $S^3$ term is inconsequential for our mechanism, because it only shifts $\vev{S}$  by an amount $O({\tilde m}^2/M_{1,2})$, but leaves the parameters $\mu$ and $B\mu$ in eqs.~(\ref{eccomu}) and (\ref{eccobmu}) unchanged.
With the introduction of an $S^3$ term in the superpotential, in the limit $M_{1,2}\to 0$ this model smoothly interpolates with the NMSSM with singlet-messenger couplings studied in ref.~\cite{sla}. Since $M_{1,2}$ determine the mass of $S$, the NMSSM contains a weak-scale singlet in the low-energy spectrum, which is absent in our model.

On the other hand, the appearence in the superpotential of a linear term in $S$ with coefficient $O(M_{1,2}^2)$ would invalidate our results. 
Indeed, since $S$ and $M_1$ must carry the same quantum numbers, a linear term $M_1M_2 S$ in the superpotential cannot be forbidden by symmetry arguments. 
Of course, non-generic superpotentials are technically natural, and the particular form of \eq{suppot1} could be the consequence of some special dynamics at the cut-off scale. Nevertheless, it is interesting to investigate if it is possible to construct models in which the form of the superpotential is dictated by symmetry. In the next section we illustrate such an example.

\section{The model}
\label{sec3}

The model involves two singlet superfields $S$, $N$ and two pairs of chiral messengers $\Phi =(\Phi_1,\Phi_2)$ and $\bar \Phi =({\bar \Phi}_1,{\bar \Phi}_2)$ with superpotential 
\beq
W=N\left( \lambda H_uH_d+\frac{\lambda_1}{2} S^2-M_S^2\right)+ \xi  S {\bar \Phi}_1\Phi_2 + X \left( {\bar \Phi}_1\Phi_1+{\bar \Phi}_2\Phi_2\right) .
\label{suppot2}
\eeq
The superpotential in \eq{suppot2} has the most general form invariant under a global $U(1)_X$ symmetry
with charges $[X]_X=1$, $[\Phi_1]_X=[{\bar \Phi}_2]_X=-1$, and an $R$-symmetry under which
$[N]_R=2$ and all messenger fields ($\Phi_i$ and $\bar{\Phi}_i$) carry charge one. Since $H_uH_d$ has zero $R$-charge, a bare superpotential $\mu$-term is forbidden. The appearance of $H_uH_d$ in the K\"ahler potential is however not constrained, thus allowing the  generation of $\mu$ once supersymmetry is broken\footnote{The situation here parallels the natural implementation of the mechanism of ref.~\cite{giumas} in supergravity. By $R$-symmetry there is no $H_uH_d$ superpotential term. However the allowed  D-term $[\phi^\dagger\phi H_u H_d]_D$, with $\phi$ the chiral compensator,
gives rise to the right $\mu$ and $B$ once $F_\phi\not=0$. 
}.  We omitted the bilinears $NS$ and $\bar{\Phi}_1 \Phi_2$ by imposing a $Z_2$ parity under which $S$, $\Phi_1$ and $\bar{\Phi}_1$ are odd. The inclusion of these terms are inconsequential for our mechanism and the $Z_2$ parity is not strictly necessary.

After integrating out the messengers at the scale $X$, we can express the kinetic term for $S$ as
\beq
K =  Z_S\left( X,X^\dagger \right) S^\dagger S, ~~~~~Z_S\left( X,X^\dagger \right)=1-\frac{5\xi^2}{16\pi^2}\ln \frac{X^\dagger X}{\Lambda^2},
\eeq
where $Z_S$ is the wave-function renormalization of $S$. The kinetic term becomes canonically normalized by redefining 
\beq
S\to Z_S^{-1/2}\left. \left( 1-\frac{\partial \ln Z_S}{\partial X} F\theta^2\right) \right|_{X=M}~S.
\eeq
The superpotential and the soft-breaking potential, below the messenger scale $M$, then become
\beq
W= N \left( \lambda H_uH_d + \frac{\lambda_1}{2} S^2 - M_S^2 \right) ,
\eeq 
\beq
V_{\rm soft} =
{\tilde m}_S^2\left| S \right|^2 +\left( A_S \lambda_1 NS^2 +{\rm h.c.}\right) ,
\eeq
where 
\beq
{\tilde m}_S^2=-\left. \frac{\partial^2 \ln Z_S}{\partial \ln X \partial \ln X^\dagger}\right|_{X=M} \frac{FF^\dagger}{MM^\dagger}, ~~~~A_S=\left. \frac{\partial \ln Z_{S}}{\partial \ln X}\right|_{X=M} \frac{F}{M} .
\label{softp}
\eeq
The soft scalar mass of $N$ can be ignored, working at the leading order in $\tilde{m}/M_S$.
The minimum of the potential is attained at 
\bea
\vev{ N} & = & -\frac{A_S^\dagger}{\lambda_1} + {\cal O} \left( \frac{{\tilde m}^2}{M_{S}}\right), \\ 
\vev{S} & = & \sqrt{\frac{2}{\lambda_1}}M_S \left( 1+
\frac{ |A_S|^2-{\tilde m}_S^2 }{ 2\lambda_1 M_S^2}\right) +O\left( \frac{{\tilde m}^3}{M_S^2}\right).
\label{vevs}
\eea
In terms of the vacuum expectation value of $N$ and $S$, we can express $\mu=\lambda \vev{N}$ and $B\mu = -\lambda \vev{F_N}$, where $F_N=-\partial W^\dagger /\partial N^\dagger$.  
As a result, we get $\mu$ and $B$ as follows,
\beq
\mu =  - \frac{\lambda}{\lambda_1} A_S^\dagger ,
\eeq
\beq
B = \frac{{\tilde m}_S^2-|A_S|^2}{A_S^\dagger}.
\eeq
The soft parameters in \eq{softp}, evaluated at a renormalization scale equal to the messenger mass $M$, are given by
\beq
{\tilde m}_{S}^2 = \xi^2 \left( 35 \xi^2 - 16g_s^2-6 g^2  -\frac{10}{3}g^{\prime 2} \right) ~\left| \frac{F}{16\pi^2 M}\right|^2,
\eeq
\beq
A_S = -  5\xi^2\left( \frac{F}{16\pi^2M} \right) .
\eeq
In terms of lagrangian parameters, $\mu$ and $B$ are expressed as
\bea
\mu & = & \frac{5 \lambda\xi^2}{\lambda_1} \left( \frac{F}{16\pi^2M} \right)^\dagger ,\\
B & = &   \left( \frac{16}{5} g_s^2+ \frac{6}{5} g^2  +\frac{2}{3} g^{\prime 2} -2\xi^2 \right) ~\left( \frac{F}{16\pi^2 M} \right) .
\label{eccomu2}
\eea

The model presented introduces no CP problem. In the low-energy lagrangian of gauge mediation, one can make all superpotential parameters real by a superfield rotation, 
leaving two possible CP invariants: ${\rm arg} (M_\lambda^* A)$ and ${\rm arg} (M_\lambda^* B)$. While $A$ vanishes at the messenger scale, the parameter $B$ has the same phase of the gaugino mass $M_\lambda$, \eq{eccomu2}, and both CP invariants are zero.
 
To summarize, the low-energy theory has the same field content of the minimal supersymmetric model with $\mu = O({\tilde m})$ generated at one loop and $B\mu= O({\tilde m}^2)$ generated at two loops. All soft terms, other than $\mu$ and $B\mu$, have exactly the usual form dictated by gauge mediation. In particular, (as opposed to the example discussed in sect.~\ref{sec2}), no new contributions to ${\tilde m}_{H_{u,d}}^2$ exist. 

The superpotential in \eq{suppot2} is very similar to that of the model in ref.~\cite{dynmu}. Nevertheless, the mechanism presented in this paper and the one of ref.~\cite{dynmu} are conceptually different, although both generate $\mu$ at one loop and $B\mu$ at two loops. One crucial difference is the presence of the $U(1)_X$ symmetry in our mechanism which dictates the form of the operator  in the K\"ahler potential, $H_uH_d \ln X^\dagger X$, as opposed to the operator $H_uH_d D^2 f(X,X^\dagger)$ of ref.~\cite{dynmu}. Because of $U(1)_X$, the $\mu$ term in our mechanism has exactly the same origin as the other soft terms of gauge mediation, {\it i.e.} the logarithmic divergence in the ultraviolet cutoff. The second important difference concerns the genericity of the superpotential. In the mechanism of ref.~\cite{dynmu}, the necessary kinetic mixing between $X$ and the singlet superfield coupled to $H_uH_d$ makes it impossible to exclude the dangerous superpotential term $XH_uH_d$ using symmetry arguments. In our mechanism, this is possible because the singlet $N$, which participates in the interaction $NH_uH_d$, is not directly coupled to the messengers. Therefore the form of the superpotential in \eq{suppot2} is the most general compatible with its symmetries. As a byproduct of the fact that $N$ is not directly coupled to messengers, we also obtain that our $\mu$-generation mechanism does not modify the usual gauge-mediation expression for the Higgs soft terms.  

\section{Conclusions}

We have presented a simple mechanism which solves the $\mu$ problem in gauge mediation. The $\mu$ term is linked to a logarithmic divergent renormalization in the K\"ahler potential. Thanks to the logarithmic dependence on the Goldstino superfield $X$, the $B\mu$ term arises only at two loops. The reason for this suppression is basically the same that forbids one-loop scalar squared masses in gauge mediation, allowing for one-loop gaugino masses and (depending on the model) trilinear couplings. New (gauge singlet) states are present with a mass, determining the scale of PQ symmetry breaking, which can be arbitrarily chosen between the weak scale and (slightly below) the messenger scale. We have focused on the case in which the new states are heavy, with an effective theory which contains only the degrees of freedom of the minimal supersymmetric model. The soft terms are exactly those of gauge mediation, with $\mu$ and $B\mu$ parametrically of the correct size. No extra contributions to the soft terms of the Higgs sector are present. There are no new CP-violating phases associated to $\mu$ or $B\mu$ and therefore the benign properties of gauge mediation with respect to flavor and CP are fully preserved. The mechanism presented here can be interpreted as a generalization to gauge mediation of the mechanism proposed in ref.~\cite{giumas}. 

We thank F.~Riva and P.~Slavich for useful discussions. 
HK was supported by the ABRL Grant No. R14-2003-012-01001-0 
and the CQUeST of Sogang University 
with grant number R11-2005-021.

\end{document}